\documentclass[thinspace,twocolumn,nofootinbib]{revtex4-2}
\usepackage{amsmath}
\usepackage{subfigure}
\usepackage{natbib}

\usepackage{graphicx}
\usepackage[a4paper, total={7in, 10in}]{geometry}
\usepackage{hyperref}
\hypersetup{
    colorlinks=true,
    linkcolor=blue,
    filecolor=magenta,      
    urlcolor=blue,
    pdftitle={Overleaf Example},
    pdfpagemode=FullScreen,
    }

\usepackage{dcolumn}
\usepackage{bm}

\usepackage{orcidlink}

\begin{document}

\title{Bridging Between Statistical Mechanics and Black Hole Evolution: \\ Theoretical Formalism}
\author{{Suvodip~Mukherjee}\orcidlink{0000-0002-3373-5236}}
\email{suvodip@tifr.res.in}
\affiliation{Department of Astronomy \& Astrophysics, Tata Institute of Fundamental Research, \\ 1, Homi Bhabha Road, Mumbai- 400005, Maharashtra, India.}
\begin{abstract}
Black holes in the Universe span nearly ten orders of magnitude in mass over a large cosmic timescale, dating back to at least the Universe's early history, when it was only a few hundred million years old. The formation and evolution of these objects remain a mystery because no direct probe can trace each black hole across the full mass range and cosmic time. This work shows, for the first time, a statistical mechanical description of black hole evolution. I show that though each black hole and its evolution are governed by stochastic processes, one can make a macroscopic description of the distribution function of black holes in the Universe, and its evolution over the entire mass range and cosmic time can be expressed in terms of the Kramers-Moyal expansion. This bridging between statistical mechanics and black hole evolution in terms of the Kramers-Moyal expansion enables the discovery of the underlying physical processes that play a role in the evolution of black holes. In the future, the application of this technique to data accessible from different messengers, such as gravitational waves and electromagnetic waves, which probe black holes in the Universe, can lead to a data-driven understanding of the underlying physical processes by this statistical mechanics technique despite the ignorance of the underlying microscopic stochastic processes. 
\end{abstract}
\date{\today}

\maketitle

\section{Introduction}
From a purely mathematical concept-- Black hole \cite{1916SPAW.......189S, PhysRevLett.14.57}, multiple observational probes have provided undeniable evidence of their existence in nature \cite{1973LicOB..24....1B, 2019ApJ...875L...6E, 1996Natur.383..415E, 1998ApJ...509..678G, LIGOScientific:2016aoc}, over a large range of masses, ranging from a few solar masses\footnote {The mass of the Sun is usually referred to as solar mass in astrophysics which is approximately $2\times 10^{30}$ Kg.} to billions of solar masses. Along with such a large dynamical range of their masses, their presence is observationally found across a large cosmic time, from a time when the Universe was about 400 million years old \cite{2023ApJ...948L..14C, Bogdan:2023ilu, Goulding:2023gqa}, to the current epoch, when it is about 13 billion years old. This observational evidence of diversity in black hole masses over at least nine orders of magnitude and their presence in the Universe over at least two orders of magnitude in cosmic time raises a natural question about their formation and evolution in the Universe, and their interplay with the growth of structure in the Universe.

In contrast to the other astrophysical systems in the Universe, such as stars (which are at the lower end of this large dynamical mass range) and galaxies (which are at the other end of the mass distribution), the inference of the physical mechanisms involved in producing and evolving a black hole's physical properties remains challenging to measure primarily for two reasons:
\begin{itemize}
    \item Apart from mass and spin,\footnote{Also, in principle, the charge of a black hole can be inferred, according to the No-Hair theorem} no other quantities are observationally measurable from a black hole. As a result, one cannot make other inferences about other physical quantities which can be useful in inferring the possible history of its formation. 
    \item In contrast to the well-understood evolution of stars and its death for different masses, the evolution of black holes in cosmic time can be governed by several mechanisms, and as a result, once a black hole is formed, it can grow and become heavier with time by several mechanisms and would remain in the Universe for the entire age of the Universe, above a mass value\footnote{A mass value above which evaporation of a black hole using Hawking radiation will not be possible}.  
\end{itemize}
As a result, from an observable quantity such as the mass (or spin) of a black hole across a large cosmic time, deciphering the underlying physics becomes extremely challenging due to the possibility of many physical mechanisms in play. The challenge in accurately inferring the formation and evolution history of black holes arises from several possible possibilities for how they grow; however, this cannot be directly inferred from observation for the entire black hole mass range and over the entire cosmic time. However, their presence cannot be ignored in the Universe, as they influence several important phenomena, including the growth of galaxies and cosmic structures in the Universe. 


In this work, I show the correspondence that can be drawn between black hole evolution and statistical mechanics, which is a branch of physics that studies the macroscopic state of a system in terms of its microstates. This work proposes analogous microstates and macrostates for black holes in the Universe and shows how their evolution can be understood in terms of non-equilibrium statistical mechanics. The work demonstrates the key physical equation that can be used to understand the measurable quantities from macroscopic states and how it can help in understanding some of the open questions on black hole evolution and growth in the Universe. 

The paper is organized as follows: in section \ref{sec-2}, we discuss the open questions on black hole evolution and why understanding it through statistical physics is appropriate. Then, in section \ref{sec-3} and \ref{sec-4}, we discuss the mathematical framework and its connection with possible observables. Finally, in section \ref{sec-6}, we discuss the conclusion and future outlook. 

\section{Motivation}\label{sec-2}
The growth and evolution of black holes in the Universe remain unknown until now. Apart from the measurement uncertainties, there remain large theoretical or modeling uncertainties, which make it even more challenging to understand their growth and evolution. Broadly, the modelling uncertainties can be classified into two parts: (i) \textbf{known unknowns:} these parts cover sectors where we know about the range of possible mechanisms and how they can play a role \cite{2010A&ARv..18..279V, Kulier:2013gda, Volonteri:2025iit, Mapelli:2021syv,2026ApJ..1000L..21K}. However, the exact physical process remains observationally unverified. As a result, certain modeling uncertainties remain, which include accretion mechanisms for different masses , hierarchical mergers of black holes and their dependence on the environment, interplay between accretion and merger, role in initial mass distribution, and its metallicity dependence, etc. (ii) \textbf{unknown unknowns:} this part covers sectors where we still do not know how they vary and impact black hole evolution. 

Along with the modeling uncertainty, there is also an intrinsic stochasticity in the growth and evolution of black holes. All black holes (even of the same mass) are not going to be impacted by the exact same procedure for accretion and merger(s) over cosmic time. Moreover, these evolution processes can change over cosmic time and will depend on the mass of the black hole. As a result, the evolution of the physical properties of black holes undergoes a stochastic process, and those stochastic processes remain vastly obscured observationally. Also, as the black hole grows with time, the evolution of black holes in the Universe over cosmic time is a non-equilibrium stochastic process due to the fact that lighter black holes can become heavier with cosmic time, but heavier ones cannot become lighter with cosmic time\footnote{Assuming there is no measurable mass loss for these black holes in the age of the Universe.}. The nature of the non-equilibrium process depends on the timescale of different physical processes in play and their comparison to the age of the Universe. Processes that occur faster than the age of the Universe will dictate the evolution of the black hole properties and hence will drive the non-equilibrium process. Several known astrophysical processes, such as formation of black holes from stars, accretion rate of matter onto a black hole, and merger rate of black holes with other black holes (or stars), are on a timescale shorter than the age of the Universe, as inferred from several observations. As a result, the non-equilibrium stochastic processes remain a valid terminology for most of the processes.  

Though the underlying processes can be non-equilibrium and stochastic, there remain measurable physical properties of a black hole which can be inferred from observations at different cosmic redshifts (and hence at different cosmic times). Such observations are possible from different probes such as X-ray binaries, AGNs, quasars, gravitationally lensed systems, black hole imaging, stellar kinematics, velocity dispersion of galaxies, and gravitational wave observations. These different probes make it possible to determine properties such as the mass (or spin) of the black hole and the redshift (which is connected to the cosmic time for a given cosmological model) at which they are detected. As a result, these physical properties can be inferred in a deterministic fashion from astrophysical observations within some measurement uncertainties. Though the exact underlying physical processes that contributed to producing a black hole remain stochastic, their deterministic properties such as mass, spin, redshift (or distance) can still be measured. I define these deterministically observable physical properties of black holes as \textbf{Macrostates}, and the number density of the macrostates (say mass $m$) at a fixed cosmic time $\tau$ is defined as the distribution function $\rho_{m, \tau}$.  

\begin{figure}
    \centering
    \includegraphics[width=1.\linewidth]{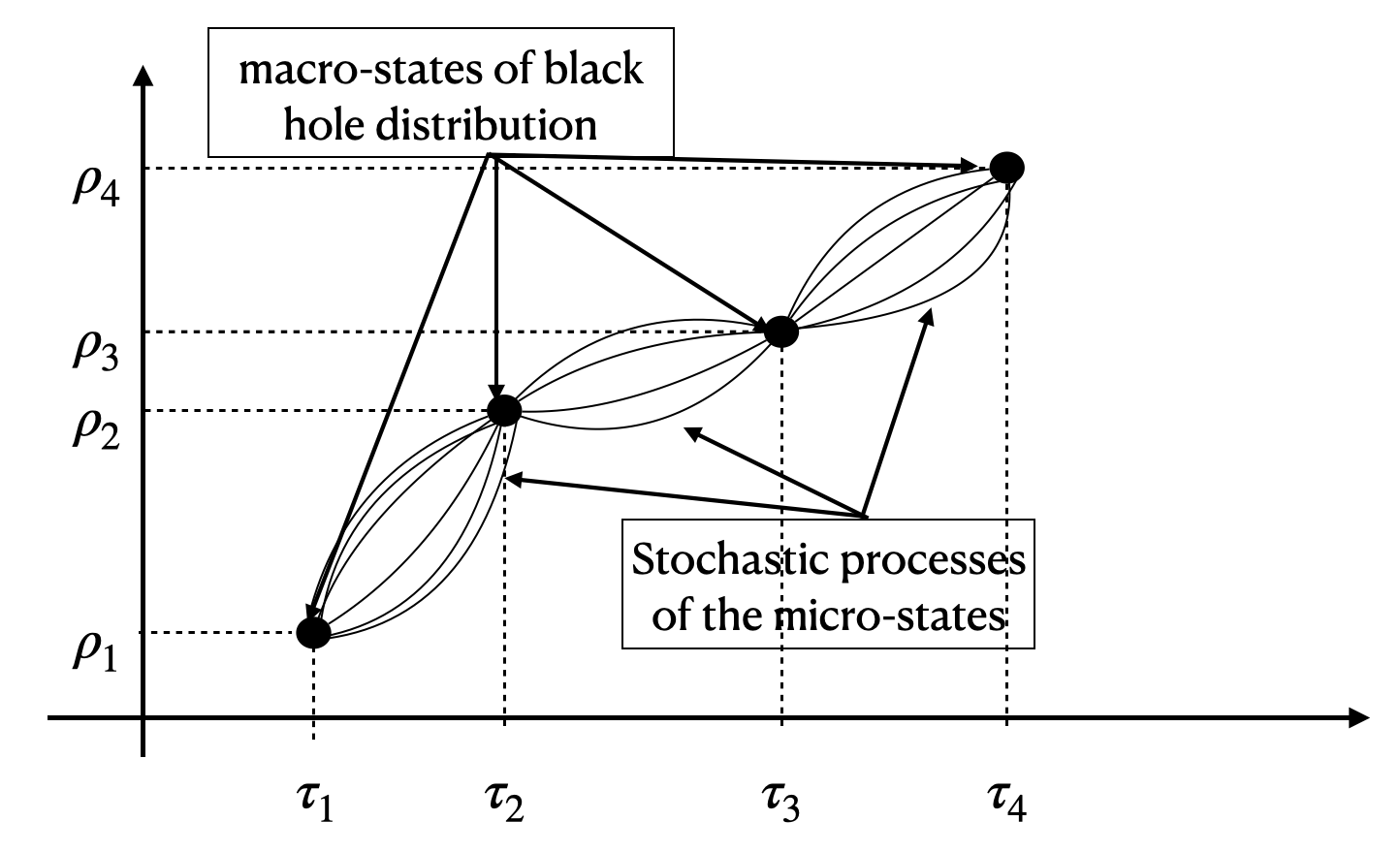}
    \caption{We show here a schematic diagram which captures the correspondence between the microstates and macro-states of black hole evolution. The big black dots denote the distribution function of the black holes $\rho_{m, \tau}$ with cosmic time $\tau$, and the lines denote an ensemble of stochastic processes of the microstates which is driving the evolution.}
    \label{fig:path}
\end{figure}
Though the stochastic processes remain indeterminate, the distribution function and its cosmic evolution with time can be useful in unveiling the physical properties and determining the properties of the equation of motion of the distribution function of black holes over cosmic time. This aspect seeds the concept behind the study of black holes in the Universe using non-equilibrium statistical mechanics.  The non-equilibrium statistical mechanics description of black holes in the Universe will capture all the measurable properties that govern the evolution of the distribution function and make a correspondence between the two areas of physics: statistical mechanics and astrophysics of black holes.  In Fig. \ref{fig:path}, we show a schematic diagram behind the concept of connecting the macrostates and microstates. In this diagram, the evolution of the macrostates, which is the black hole distribution function $\rho_{m, \tau}$ (denoted by big black dots), is a composition of all black holes in the Universe at that epoch (denoted by $\tau$). The lines denote the possible ways the macrosates can evolve from one point of the distribution function to another with time. Each line is composed of multiple stochastic processes of the microstates that govern the evolution of the measurable macrostates. At present, these transitions are unknown from one point of cosmic time to another in the distribution function $\rho_{m, \tau}$. The statistical mechanical description gives a mathematical framework to understand the physical process of the evolution of the distribution function.

\section{Correspondence between Statistical Mechanics and Black Hole Evolution}\label{sec-3}
In this section, I describe the mathematical framework to describe the correspondence between non-equilibrium statistical physics and black hole evolution. As shown in Fig. \ref{fig:path}, we are interested in understanding the evolution of the distribution function of black holes from one cosmic time to the other. In this setup, the number of black holes detected at a particular cosmic time ($\tau$) with a physical property (say, the mass of a black hole\footnote{In the remainder of this paper, we will focus on the mass of the black hole as a physical property. However, the formalism is applicable also for other intrinsic properties of black holes such as their spin.}), divided by the total number of black holes detected across the entire range of the parameter space. We define this quantity as the black hole distribution function $\rho(m, \tau)$. Then each black hole contributing to the distribution undergoes non-equilibrium stochastic processes along its entire trajectory before reaching that mass. For example, the black hole could have formed from a star directly without any further gain in mass, or could have gained mass after forming a black hole in its past, before getting detected, or could have been an outcome of mergers from previous black holes (or any other compact objects) in the past, or direct collapse from initial perturbations for primordial black holes, or some other processes. Though none of these steps would be known, the physically measurable quantity is the distribution function, and how the distribution function evolves with time is governed by the time-dependence of the underlying stochastic processes in play. The only physically meaningful quantity to infer from observation will be based on the properties of the distribution function of the macro-states. 

For this, we set up the following principles: 
\begin{itemize}
    \item At any redshift (or cosmic time) of interest, the black hole distribution does not change in real time, and it appears stationary within the human observation time. In other words, we are not observing the Universe at any special time, and the black hole distribution that we detect is a representative distribution at that redshift (or cosmic time). 
    \item Between two cosmic epochs or redshifts, the distribution function of black holes can evolve with time. 
    \item Any particular black hole contributing to the distribution function of black holes is an outcome of numerous stochastic physical processes, and we do not have any probe to explore these processes; the only physically measurable meaningful quantity is the distribution function of black holes. 
\end{itemize}

\begin{figure}
    \centering
    \includegraphics[width=1\linewidth]{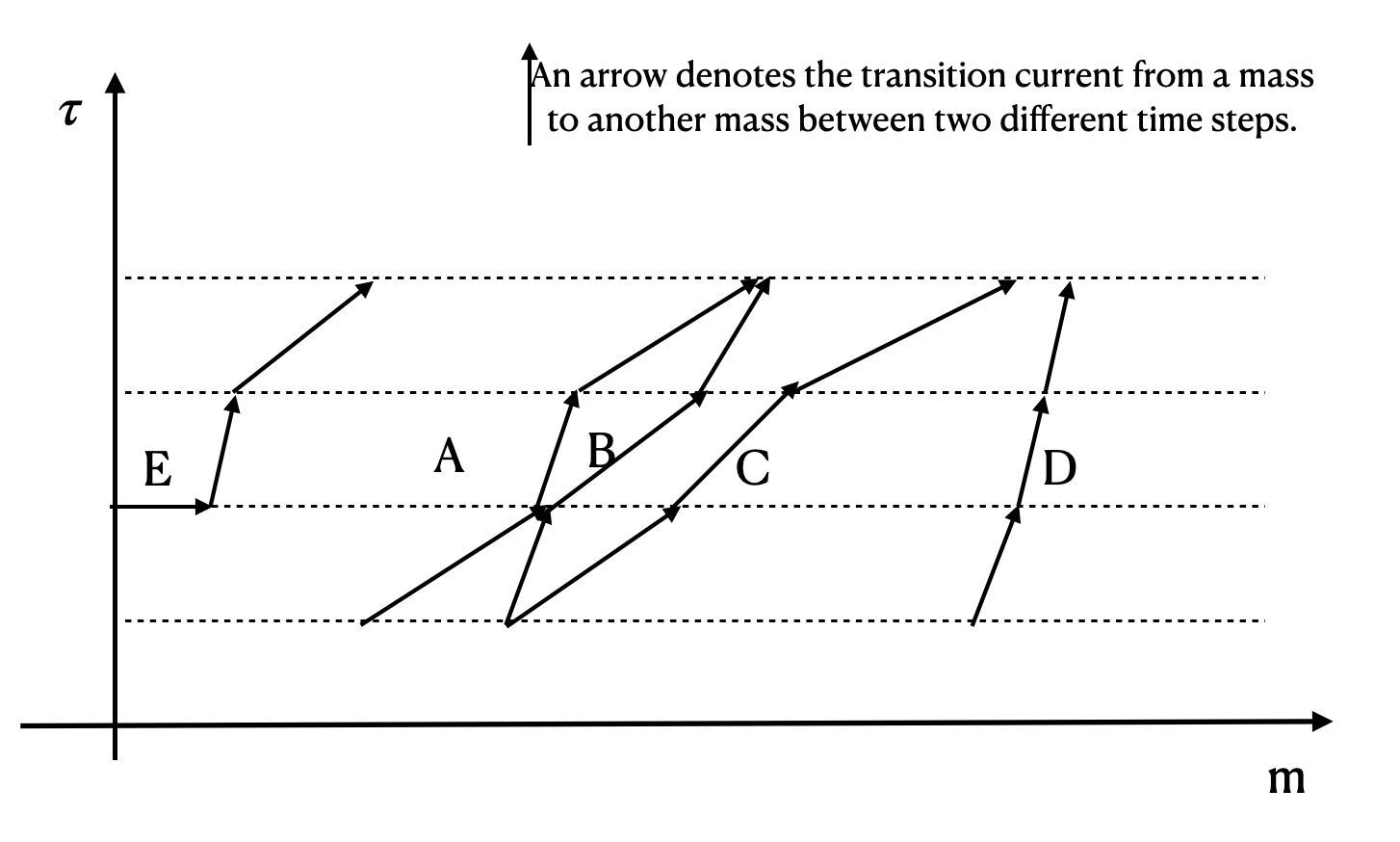}
    \caption{This plot shows the mass and cosmic time diagram to show the black hole evolution from one mass to another for different trajectories labeled from A to E. The arrows denote the transition current from one mass to another, which is governed by different stochastic processes. The ensemble of these different transition currents governs the macroscopic evolution over cosmic time.}
    \label{fig:current}
\end{figure}
With these principles of the method, let's define the true state distribution function of black holes $\rho_{m, \tau}$, where the density labels are the mass of a black hole $m$ at a cosmic time $\tau$ as shown in Fig. \ref{fig:current}. This diagram shows that each black hole of a certain mass $m$ and at a cosmic time $\tau$ undergoes a transition to another mass, always higher than the initial mass, through different processes, which are denoted by an arrow. For different trajectories such as A, B,C and D, the rate of evolution between two points can be different and can depend on the masses and time. For trajectory E, the first part of the trajectory, denoted by a horizontal arrow, denotes the formation of a black hole with that mass at that time, which again starts evolving with time. The ensemble of all these transition currents decides the net evolution of the macrostates, and from which one can draw conclusions about the underlying physical processes. With this picture in mind, one can develop a statistical mechanics description of the black hole evolution by considering all possible contributions to this true state distribution function from all other masses from a previous time $\tau'$ within an infinitesimal time window $\Delta \tau= \tau -\tau'$ as
\begin{align}
    \rho_{m,\tau}=& \rho^s_{m, \tau} + \bigg[\sum_{m'\leq m, \tau'}\mathcal{J}_{(m, \tau);(m', \tau')} \rho_{m',\tau'} \\\nonumber & - \sum_{m\leq m', \tau'}\mathcal{J}_{(m', \tau);(m, \tau')} \rho_{m,\tau'} \bigg]\frac{\partial \tau}{\partial z}(z)dz,
\end{align}
where $\rho^s_{m, \tau}$ denotes the number density of black holes formed at that cosmic time $\tau$ within a time window $\Delta \tau \equiv \frac{\partial \tau}{\partial z}dz$, which corresponds to a cosmic redshift $z$ and redshift window $dz$. The second quantity determines the number of black holes of mass $m'$ which formed at an earlier time $\tau'$ and evolved by various processes in the past until cosmic time $\tau$. As a result, this can be considered a transition current for black holes from $m'$ at $\tau'$ that can become a black hole of mass $m$ at time $\tau$. The causality condition, and assuming that a black hole in the astrophysical mass range only grows in time\footnote{Ignoring any substantial contribution from Hawking radiation}, will require $\tau' < \tau$ and $m'< m$. The third term captures the loss of black holes from a mass $m$ to higher masses. If we define, the source term as $\mathcal{\bar J}_{(m, \tau);(\{0\}, \tau')} \equiv \rho^s_{m, \tau}$, then the above equation can be written 
in a compressed form as
\begin{align}
    \rho_{m,\tau}=&  \bigg[\sum_{\{0\} \leq m'\leq m, \tau'\leq \tau}\mathcal{\bar J}_{(m, \tau);(m', \tau')} \rho_{m',\tau'} \\ \nonumber &- \sum_{\{0\} \leq m\leq m', \tau'\leq \tau}\mathcal{\bar J}_{(m', \tau);(m, \tau')} \rho_{m,\tau'}\bigg]\frac{\Delta \tau}{\Delta z}(z)dz,
\end{align}
where $\mathcal{\bar J}_{(m, \tau);(m', \tau')}$ includes the transition probability from a black hole to a black hole and also the transition probability from a non-black hole to a black hole (which was previously captured in terms of the source term $\rho^s_{m, \tau}$). An important physical property of this transition current, which we impose, is that $\mathcal{J}_{(m, \tau);(m', \tau')}=0$ for $m'>m$, as black holes do not decay (ignoring Hawking radiation) with time, and hence the probability of a black hole from higher mass to lower mass is zero. Also, due to the first point in the principles written above, that we are not observing the Universe at any special time, the distribution function of the macrostates is stationary in our observation time. Though the distribution function evolves over cosmic timescales, within our observation time the distribution at any fixed cosmic epoch $\tau$ will appear stationary. As a result, we will describe this distribution function as a locally-stationary distribution function of a non-equilibrium stochastic process. This transition current captures the underlying physical mechanisms that can make a black hole form and grow. Though the underlying physical mechanisms are stochastic (and hence uniquely and accurately capturing all different contributions to the transition current is challenging), some key physical aspects about this transition current can be inferred from the macro-state distribution functions. An important physical property to note is that, for certain processes, one can write 
$\mathcal{J}_{(m, \tau);(m', \tau')}$ using the Chapman-Kolmogorov \cite{Chapman1928, Kolmogorov1931} equation as 
\begin{equation}
    \mathcal{J}_{(m, \tau);(m', \tau')}= \int \mathcal{J}_{(m, \tau);(m'', \tau'')} \mathcal{J}_{(m'', \tau'');(m', \tau')} dm''.
\end{equation}
This implies that the transition current between $(m, \tau)$ and $(m', \tau')$ will be the same as the product of the transition currents between $(m, \tau)$ and $(m'', \tau'')$ and between $(m'', \tau'')$ and $(m', \tau')$, integrating over all the masses it can take\footnote{Remember that the condition that the transition current is non-zero only for the transition from lower mass to higher mass is also satisfied by this equation.}. This equation is physically correct for all the processes which depend only on the initial and the final stage, and not the intermediate steps in between. But it is important to note that this will not be true for all astrophysical processes, such as hierarchical mergers of black holes by emitting gravitational waves, as the transition currents from two different masses at two different epochs would not only depend on the initial and final state, but also on the intermediate exact processes in between, because the amount of mass loss in GW during a merger depends on other source properties apart from only the masses of the individual black holes. 

As a result, between two macrostates observed at two different cosmic epochs, the evolution history will not only depend on the previous time-step, but on the entire history in time. Hence, the process will not be a Markovian process in reality. The mathematical framework to extract the physically meaningful quantities from the macrostate distribution of black holes can be constructed in terms of the 
stochastic differential equation can be written as

\begin{align}\label{eq:master}
\frac{\partial \rho_{m, \tau}}{\partial \tau}
&= \int d\Delta\tau \bigg[\int_{m'< m} dm' \mathcal{\bar J}(m' \rightarrow m, \tau -\Delta\tau) \rho_{m', \tau-\Delta\tau} \bigg] \\ \nonumber & - \bigg[\int_{m''>m} dm'' \mathcal{\bar J} (m\rightarrow m'', \tau -\Delta\tau) \rho_{m, \tau-\Delta\tau}\bigg], 
\end{align}
where, in the first line of the above equation, we have written a balance equation in terms of the transition currents (denoted by the notation $\mathcal{\bar J}(m_A \rightarrow m_B, \tau -\Delta\tau) \equiv \mathcal{\bar J}_{(m', \tau);(m, \tau-\Delta \tau)}$)  for the mass states $m$ to be populated from a lower mass state $m'$ and the mass state to be depleted due to transition from the mass state $m$ to higher masses $m''$. The integration window in time depends on the past time up to $\tau$, where the lower limit depends on how far back in time one is aware of the macrostate distribution function of black holes. The lower limit of the integration can be chosen depending on the observable, and in principle it can be taken as $\tau''\equiv \tau-\Delta\tau\sim0$, which implies being close to the epoch at the beginning of the Universe. Since the micro-state processes that populate the masses $m$ from $m'$ depend on the availability of sources in $m'$, the probability of transition depends on the distribution function in $m'$, and similarly the probability of transition depends on $m$, to estimate the number of which can leave this mass bin, at an earlier time. It is very important to note here that, on cosmological scales, objects detected at larger redshift connect with an earlier epoch of the Universe; as a result, the dependence of the change in the distribution function will depend on the distribution function at an earlier epoch, and hence at a higher redshift. 

If we define the characteristic function $\mathcal{P}(k, m, \tau, \Delta \tau)$ in terms of the moments of the distribution function $M_n(m, \tau, \Delta \tau)$ as
\begin{align}
    \mathcal{P}(k, m, \tau, \Delta \tau)
    &\equiv 1+ \sum_{n=1}^{\infty} (ik)^n \frac{M_n(m, \tau, \Delta \tau)}{n!},
\end{align}
then the corresponding probability distribution is connected to the moments by \footnote{Using the relation between characteristic function and probability distribution function through Fourier transformation.} 
\begin{widetext}
\begin{align}
\mathcal{\bar J}(m' \rightarrow m, \tau -\Delta\tau) &=  \frac{1}{2\pi}\int_{-\infty}^{\infty} e^{-ik(m'-m)} \mathcal{P}(k, m, \tau, \Delta \tau)dk\nonumber,\\
&= \frac{1}{2\pi}\int_{-\infty}^{\infty} e^{-ik(m'-m)} \bigg[1+ \sum_{n=1}^{\infty} (ik)^n \frac{M_n(m, \tau, \Delta \tau)}{n!} \bigg] dk.
\end{align}
\end{widetext}
Using the relation that 
\begin{equation}
    \frac{1}{2\pi} \int_{-\infty}^{\infty}  (ik)^n e^{-ik(m'-m)} dk= \bigg(\frac{-\partial}{\partial m'}\bigg)^n\delta(m'-m),
\end{equation}
we can write down the transition probability in terms of the moments as
\begin{align}
  \mathcal{\bar J}(m' \rightarrow m, \tau -\Delta\tau) & = \delta (m'-m) \\ \nonumber  + & \delta (m'-m) \sum_{n=1}^{\infty}  \bigg(\frac{-\partial}{\partial m'}\bigg)^n\frac{M_n(m, \tau, \Delta \tau)}{n!}. 
\end{align}
If we expand the moments in terms of the Taylor series expansion of $\Delta \tau$ as 
\begin{equation}\label{eq:taylor}
    \frac{M_n(m, \tau, \Delta \tau)}{n!}= D^n(m, \tau)\Delta \tau+ \mathcal{O}\{(\Delta \tau)^2\},
\end{equation}
then the above equation for the evolution of the probability distribution function can be written as 
\begin{equation}\label{eq:master2}
\frac{\partial \rho_{m, \tau}}{\partial \tau}
=\sum_{n=1}^{N}\bigg(-\frac{\partial}{\partial m}\bigg)^n D^n(m, \tau)\rho_{m, \tau}.
\end{equation}
The equation in terms of the coefficients $D^n(m, \tau)$ and their derivatives with respect to mass is a result of the perturbative expansion in $\Delta \tau$, which can capture the evolution of the distribution in terms of the underlying physically interpretable properties. An equivalent expression in statistical physics is called the Kramers-Moyal expansion \cite{Moyal1949StochasticPA, 1940Phy.....7..284K,risken1989fpe}, and the coefficients are called the Kramers-Moyal coefficients. As cosmic time $\tau$ and redshift $z$ are uniquely connected in a cosmological model as $\Delta \tau= (\partial \tau/\partial z)\Delta z$, we can write down the above equation in redshift as
\begin{equation}\label{eq:master2}
\frac{\partial \rho_{m, z}}{\partial z}
=\sum_{n=1}^{N}\bigg(-\frac{\partial}{\partial m}\bigg)^n D^n(m, z)\rho_{m, z}. 
\end{equation}

The coefficients of the Kramers-Moyal expansion capture several important behaviors of the distribution function. As we are dealing with a physical system under a stochastic process, and the evolution of the distribution depends on a stochastic differential equation, the values of these terms would be non-zero. The term with $n=1$ captures the drift in the distribution function. Similarly, the term for $n=2$ captures the effect of diffusion on the distribution function. The terms with $n\geq 3$ can also be non-zero for the black hole distribution function, capturing the non-Gaussian nature of the distribution function. However, following the Pawula theorem \cite{Pawula1967Generalizations}, if the series do not truncate after $n=2$, then it will have all the higher-order terms, for any transition probability which is non-negative. It is important to note here that these coefficients can depend on the cosmic time and can evolve with cosmic time. The measurement of these quantities will provide the measurable physical description of the distribution function and its time evolution over the entire mass range. The connection of the above equation with observations will be discussed in the next section. 

The measurement of the coefficients of the Kramers-Moyal expansion and their evolution with cosmic time will shed light on some of the key aspects of whether the formation of black holes in the Universe is a drift-dominated process and/or a diffusion-dominated process, and what the typical strengths of these terms are. It would also reveal the skewness, kurtosis, and other higher-order terms in the Universe. A major advantage of bridging statistical mechanics with black hole evolution is that it helps in revealing different physical processes and their corresponding redshift dependence. The inference of these coefficients over cosmic time in the framework of statistical mechanics would allow us to also estimate the distribution function of black holes in the future in a perturbative limit for the $\Delta \tau/\tau <1$ case using the evolution equation given in Eq. \eqref{eq:master2}. 

\section{Applicability of the framework for astrophysical observations}\label{sec-4}
The solution of the evolution of the distribution function in terms of the coefficients of the Kramers-Moyal expansion is an important step to understand the underlying physical processes. Connecting this equation with observables would make it possible to understand the \textit{measurable} physical quantities from macrostates. There are broadly two approaches to this: (i) forward-modeling by capturing different physical mechanisms and calculating the terms in the Kramers-Moyal expansion, which can be matched with observations, (ii) backward-modelling by reconstruction of the coefficients in the Kramers-Moyal expansion from different observations. 

The first option of forward-modelling is challenging because of the stochasticity in the known processes and the associated known unknowns and unknown unknowns. Whereas the second option is a direct inference (or reconstruction) of the terms from the data and avoids model assumptions connected to the underlying stochastic processes. This direct inference from data makes it possible to extract the information content in a model-independent way.  The inference of the coefficients in the Kramers-Moyal expansion is possible under two conditions: (i) the masses are inferred over a large range, in order to estimate the coefficients, and (ii) the measurement of the black hole masses is possible over a wide redshift range in a tomographic fashion such that tomographic width in cosmic epoch $\Delta \tau$ is much less than the age of the Universe at that redshift $\tau(z)$, such that $\Delta \tau/\tau(z) <<1$ is satisfied, and hence the approximation in Eq. \eqref{eq:taylor} is valid. This will become feasible in the coming years with the aid of multiple observations covering different observational probes using different messengers, starting from multi-band electromagnetic observations covering radio \cite{2019ApJ...875L...6E} to gamma-rays \cite{1999ApJ...518..356P}, and multi-band GW observations covering from nano-hertz \cite{Sesana:2025udx} to kilo-hertz \cite{Punturo:2010zz, Reitze:2019iox} in the frequency range. These observational probes make it possible to access black holes over nearly ten decades in magnitude across cosmic time.  As a result, from the combination of different probes of black holes using multiple cosmic messengers, a comprehensive picture of the macroscopic black hole distribution and its connection with the underlying physical processes can be captured in terms of the coefficients of the Kramers-Moyal expansion, and their dependence on redshift. In future work, we will develop a statistical estimator which can be used to infer the coefficients of the Kramers-Moyal expansion from multi-messenger observations. 

\section{Conclusion and Future Outlook}\label{sec-6}
In this work, we demonstrate for the first time the correspondence between statistical mechanics and black hole evolution in the Universe. We show that though numerous known and unknown processes can drive the formation and evolution of the Universe, which are not deterministically plausible to be inferred from any current techniques, one can make a connection to the macroscopic state and understand 
the microscopic state behavior of the black holes in the Universe in the framework of statistical mechanics. We show that the underlying physical processes driving such effects of individual black holes can be considered as microscopic processes, which lead to an observable impact on the black hole distribution function, which is the macroscopic state of the black holes. The black hole distribution function can be constructed in terms of the Kramers-Moyal expansion, and hence we can build an understanding of different physical mechanisms that are in play in driving the evolution of the black holes in the Universe.    

We derive the correspondence between statistical physics and black hole evolution, and show that one can express the cosmic evolution of black holes in terms of the Kramers-Moyal expansion. The coefficients $\mathcal{D}^n$ of the Kramers-Moyal expansion are connected to physical processes such as drift and diffusion for $n=1$ and $n=2$, respectively, and the time (or redshift) evolution of these quantities over the mass range of black holes in the Universe. 

In the future, this particular proposed connection between black hole evolution and statistical physics can be explored using observations by combining different tracers of black holes using different cosmic messengers, such as gravitational wave signals and electromagnetic wave signals. On the gravitational wave side, masses of the individual compact objects can be inferred from a few solar masses to millions (or billions) of solar masses using signals from hecto-hertz to nano-hertz, respectively. On the other hand, with electromagnetic signals covering multiple bands, we can make direct or indirect inferences of black hole masses from a few solar masses using sources such as X-ray binaries to billions of solar masses from observations of objects such as AGNs. In future work, we will develop an estimator that can combine different multi-messenger tracers to black holes across cosmic time. 

The correspondence between statistical mechanics and the cosmic evolution of black holes demonstrated in this work provides the theoretical platform that can capture the measurable effects at the macroscopic state of the black hole distribution function in terms of underlying physical mechanisms involved in the evolution of the black hole distribution (which are related to the coefficients of the Kramers-Moyal expansion), and their evolution with cosmic time. In the future, the measurement of these quantities by combining observations from both GW and EM sectors will open a completely new paradigm to understand the growth of compact objects across mass scales and cosmic time, and will be able to bring new insights to the cosmic evolution of black holes in the Universe and will be able to shed light on one of the open questions in astrophysics related to how black holes form at earlier times. 

\section*{Acknowledgments}
The author acknowledges support from the ICTP through the Associates Program (2026-2031), where most of the work was carried out. The author is very thankful to ICTP for providing support and hospitality for carrying out this research. This work is a part of the ⟨Data$|$Theory⟩ Universe Lab, which is supported by the Tata Institute of Fundamental Research (TIFR) and the Department of Atomic Energy, Government of India. We acknowledge the support of the Department of Atomic Energy, Government of India, under Project Identification No. RTI 4012. This research is also supported by the Prime Minister Early Career Research Award, Anusandhan National Research Foundation, Government of India.

\bibliographystyle{apsrev4-1}
\bibliography{main}

\end{document}